\documentclass[a4paper,showkeys,floatfix,aps,pre,reprint,groupedaddress]{revtex4-1}
\usepackage{amsmath,amssymb}
\usepackage{graphics,graphicx}
\usepackage{dcolumn,bm}
\usepackage{psfrag}
\usepackage{color}
\newcommand{\cu}
{\affiliation{Department of Physics, University of Calcutta,
92 Acharya Prafulla Chandra Road, Kolkata 700009, India.}}

\newcommand{\germany}
{\affiliation{Max Planck Institute for Dynamics and Self-Organization, Am Fa{\ss}berg 17, 37077 G\"{o}ttingen, Germany}}

\begin{document}
\title{Critical noise can make the minority candidate win: The US Presidential Election cases}

\author{Soumyajyoti Biswas}

\email{soumyajyoti.biswas@ds.mpg.de}
\germany
\author{Parongama Sen}

\email{psphy@caluniv.ac.in}
\cu
\begin{abstract}
A national voting population, when segmented into groups like, for example, different states, can yield a
counter-intuitive scenario where the winner may not necessarily get the most number of total votes. A recent example
is the 2016 presidential election in the US. We model the situation by using interacting opinion dynamics models 
and look at the effect of coarse graining near the critical points where the spatial fluctuations are high.
We establish that the sole effect of coarse graining, which mimics the `winner takes all' electoral college system in the US,
can give rise to finite probabilities of
such events of minority candidate winning even in the large size limit near the critical point. 
The overall probabilities of victory of the minority candidate can be predicted from the models which indicates that 
one may expect more  instances of minority candidate winning in the future.

\end{abstract}


\maketitle
\section{Introduction}
The 2016 presidential election in the US is one of the rare cases where the winner did not get the most number of
total votes. This follows from the `winner takes all' rule placed at individual states (with some exceptions) and
the relatively low margin of difference in votes between the major candidates. In the past, there have been substantial progress
in the efforts to develop mathematical models  of opinion dynamics within an interacting society (see e.g., 
\cite{book1,rmp,galam_book,socio_book,travieso,fort-cast07,alves}) 
 with analysis of relevant data (see e.g., \cite{chatt_elec,sinha}).  The formulation, sustenance
or change of the individual opinions construct an example of a complex system, where the emergent behavior is qualitatively different 
from the individual ones. The focus has been, among others, to develop a scenario of 
phase transition to a global consensus that determines the favorite among two or more choices, for example,  the winner of an election. 
However, in the large system size limit, a state of consensus excludes the above mentioned possibility of the victory
of a global minority candidate even for a partitioned population. Nevertheless, such event is not unique in the 2016 election, but has happened 4 times (1876, 1888, 2000, 2016) \cite{note, database}
in 49 elections and twice in the past 16 years. It is therefore important to analyze the probability of such events and its
dependence on various parameters, particularly in view of its growing occurrence and relative unpredictability.

\begin{figure}
\includegraphics[width=8cm]{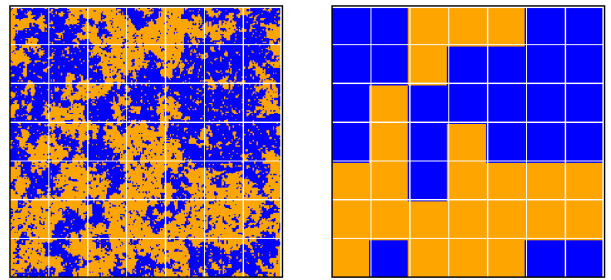}
\caption{The snapshot of the spatial organizations of the opinion values in an Ising model before (left) and after (right) the coarse-graining.
Although the more spins are up (orange or light-grey, 22382) than down (blue or dark-grey, 21718) in the original system, due to the strong spatial correlations, 
in the coarse grained system the
down spins forms  the majority (25-24). The coarse graining boxes are also shown, but they do not affect the interactions between, say, 
agents on the boundary of one box with that in a neighboring one.}
\label{snapshot}
\end{figure}

The US presidential election system resembles a step of coarse graining mechanism \cite{skma}. Every state has a fixed number of electoral college votes all of whom are 
generally assigned to the candidate winning most votes in that state. The events of a candidate winning the electoral college but not the 
popular vote can therefore arise when a spatial correlation drives strong fluctuations in winning margins in different states. 
In this paper, we wish to investigate the
sole effect of coarse graining in  interactive  opinion models. They mimic the   US election process  particularly near the critical point where
no opinion has a dominant majority and the fluctuations are high. Coarse graining near the critical point can lead to reversal of sign of the 
majority before and after the step (see Fig. \ref{snapshot}).
 The aim is to arrive at a universal characterization of the probability of
a minority candidate winning  and its finite size scalings.
Finally, we compare the predictions from the model with the real instances of minority winning situations.

Previous works along this line include
random partitioning of population and conversion of the opinions of all agents to that of the 
majority \cite{galam86, galam2002,krapv,nard}, leading to an initial minority to win, which is a 
different case from this study. Here we do not convert the opinion values but look for just the effect of coarse graining.
A few other attempts had also been made prior to the 2016 election, to obtain the probability of a minority 
win in a large number of elections \cite{x1,x2,gracia}, mean majority deficit with equipopulous units (states) \cite{r1, r2},
and also the effect of prejudice among likely voters \cite{galamx3}. 

\begin{figure}
\includegraphics[width=9cm]{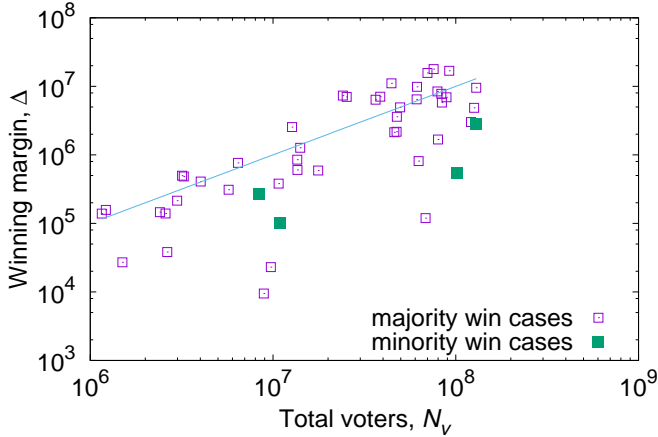}
\caption{The margins of win as a function of the total number of votes cast for the two major parties in the US presidential elections. The minority 
win cases are shown in different symbols.
The solid line is the linear relationship $\Delta \propto N_v$. The linear tendency for the relatively higher margin wins is expected 
from the models and is explained in the text later on.}
\label{US_alldata} 
\end{figure}
The following general scheme for studying  the minority winning probability is used here: 
Let $\Delta$ be the difference between the votes received by the  two highest ranked candidates while $N_v$ is the total number of voters who voted
for those two candidates (this is, in general, different from the total number of eligible voters $N$, a large fraction of which do not vote or
vote for other parties).  
One can  estimate  the probability  $w(\Delta)$ for the victory of the candidate with less number of popular votes (minority) has won, for 
specific values of  $\Delta$.    
In the models showing a second order phase transition, above the critical point $\Delta$ fluctuates about zero  and its 
distribution is obviously Gaussian. i.e., $P(\Delta) \propto \exp(-a\Delta^2/N_v)$.
The total minority winning probability is $w_{tot}=  \int w(\Delta) P(\Delta) d\Delta$ which is  dependent on the system's  
parameters, specifically on the noise driving the system.  
Since $\Delta$ and $N_v$ values are known in the real situations, it makes sense to estimate the probability $w(\Delta)$ 
 to make direct comparisons between the simulations and real data. 

 In the case of real data, we show he behavior of $\Delta$ vs $N_v$ in Fig. \ref{US_alldata}. 
In the total eligible population of voters, a significant fraction do not vote and also
a fraction vote for other parties, which is crucial in estimation the probability $w(\Delta)$, as we will discuss later. 

\section{Models}
One needs to quantify the opinions  of the people in an agent based model and also their interactions.
 A large number of opinion dynamics  models resembling spin systems exist in literature in which 
the opinions are assumed to take discrete values. The interactions
among the agents are usually chosen to favor the tendency to align while temperature and/or noise or unfavorable
 interactions tend to destroy such a global 
alignment or 
dominant order. Below a critical value of the strength of disorder, there is a global order with one opinion dominating 
and above it opinions are fragmented. Being a critical transition, however, it  will ensure the existence of large clusters 
of correlated opinion values near the critical point.

We begin by simply assigning  values $\pm 1$ randomly to the agents with equal probability. This is just to check
the sole effect of coarse-graining in the system. Of course, a more realistic approach is to have interactions among agents. We consider two
interacting models: (i) we start by considering the simplest interacting model with $Z_2$ symmetry, i.e. the Ising model. The Ising
model was used in the past to model two competing opinions, as is the case here. It has a critical point which is exactly known and 
we will study the effect of coarse graining near the critical point. (ii) Finally, we consider a model more suited 
to model opinion formation in presence of neutral voters. For this we use the kinetic exchange model (KEM) \cite{BCS} where alongside the two
major opinion groups ($\pm 1$), there is also a neutral group (0) that accounts for third party voting or non-participation. 

In the cases where we do not have non-participation or third party voters in the model (e.g. random assignments, Ising model), the total number of voters and the
number of voters for the two major parties are the same, i.e. $N_v=N$ there. In the other cases (e.g., KEM and also the real data) $N_v<N$. 

\section{Results}
\subsection{Random assignment}
\label{random_assign}
\begin{figure}
\includegraphics[width=9cm]{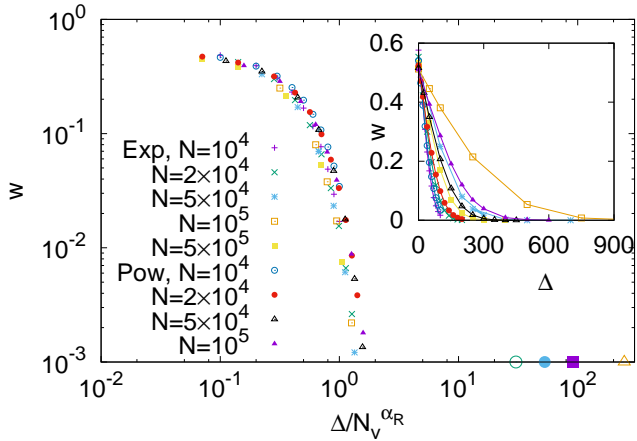}
\caption{The finite size scaling of the minority win probability ($w$) is shown for the random assignments of the probability values.
The block sizes are taken from  exponential and power law distributions (scaled in two different curves).
The scaling exponent value is $\alpha_R=0.5$. The scaled variables $r_{\alpha_R}=\Delta/N_v^{\alpha_R}$ showed by various symbols (filled square 1876, empty circle 1888, filled circle 2000 and empty triangle 2016) in the x-axis for the real data and they lie far away from
where the scaled function has become zero. The inset shows the unscaled data.}
\label{collapse_random} 
\end{figure}

As mentioned above, to capture the effect of coarse graining only, we begin by just randomly assigning opinion values ($\pm 1$) on the $N_v (=N)$ agents. 
They are then divided into $M=50$ groups and the winner in each group is assumed to have the backing of the entire group. We then calculate the 
probability $w$, i.e. the cases where the candidate winning over most number of blocks is not the candidate having more individual votes, 
 as a function of $\Delta$, which we can now fix by hand. For different system sizes, the scaling $w\sim f(\Delta/N_v^{\alpha_{R}})$ is
seen with $\alpha_{R} =0.5$ (see Fig. \ref{collapse_random}).  One can now calculate the values of $r_{\alpha_{R}} = \Delta/N_v^{0.5}$ for the instances where the minority 
candidate had turned out to be the winner in past US elections. The comparison of these points with the scaled curve shows  that 
the predicted minority win probability is practically zero. 
This result is not surprising as opinions are not in general random or uncorrelated  and clearly one needs to consider models 
where the agents have some kind of interaction.  

\subsection{Models with interactions}

\subsubsection{Ising model}
Although primarily studied as a model showing magnetic phase transition, the Ising model, can also be regarded as a binary opinion dynamics model. In fact, 
in one dimension, it is identical to the well known voter model \cite{voter}.
The thermal fluctuation here represents the  noise or  discord in the social interactions.

The Ising model is known to have an exact critical temperature $T_c =2.26918 \dots$ in two dimensions.  The state of the 
agents are    $\sigma=\pm 1$,  which corresponds to voting for  one of the two candidates. Therefore, in this case $N_v=N$. 
The agents are divided into $M=49$  groups (so that each group can be a square) resembling states of the US elections. 
 The imposition of the group identities of the individuals 
has nothing to do with
their interaction probabilities, which remain same for all. 
The groups have same number of agents but their weightage in calculating the coarse grained
values are different. This is to take into account the different numbers of electoral college votes 
from different states (population variation will give similar fluctuations).
In the steady state, the magnetization will fluctuate about a constant value; above $T_c$, this value is  equal to zero.  
One determines  the 
winner in each group by calculating the number of votes (i.e.,  the number of spins having $\sigma=+1$  and $-1$) in the group.     
The overall winner is determined by the weighted average of the winning states for each of the two candidates and not from 
the individual opinions.

\begin{figure}
\includegraphics[width=8cm]{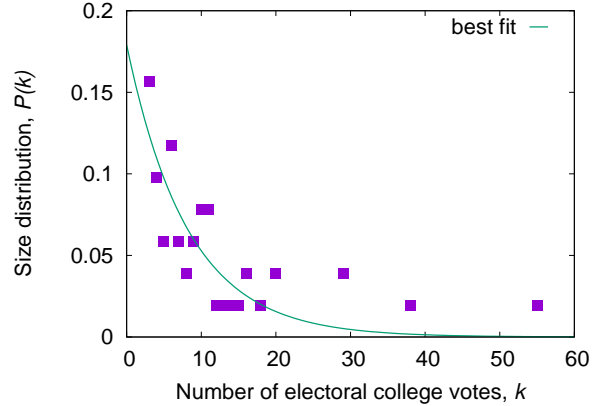}
\caption{The actual sizes of the electoral college votes $k$ and their probability of occurrence $P(k)$ in the US.
The best fit with an exponential function is shown, although in the simulations the actual values were used and not the fit.}
\label{elec_clg} 
\end{figure}
 Let the weighted sizes of the 
individual groups be denoted by $C_I=k_IS$ (with $S=N/M$ is the group size and $k_I$ denotes the weight of the group). Then we look for the quantity 
\begin{equation}
Q(t)=\sum\limits_{I\in\{up\}}C_I-\sum\limits_{I\in\{down\}}C_I,
\end{equation}
 where  $\{up\}$ is the set 
of groups in which $+1$ opinion is dominant and in $\{down\}$, $-1$ opinion is dominant. The values of $k_I$ are chosen from a distribution. 
The actual form of distribution does not influence the scaling exponents. We have considered a uniform distribution in $[0:1]$ and also
the actual sizes of the electoral college votes from different states in the US (see Fig. \ref{elec_clg}), which best matches with an exponential distribution. 
 Finally, the probability ($w$) of the event where the candidate with
less number of total vote wins, is defined as the fraction of times when $Q(t)m(t)<0$, with  $m(t)=\sum\limits_i\sigma_i(t)$ calculated in the steady state of the model.  

\begin{figure}
\includegraphics[width=8cm]{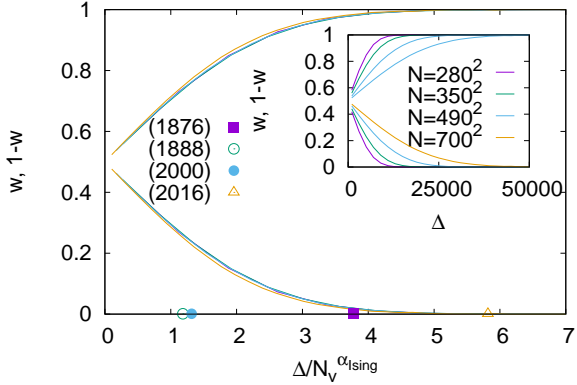}
\caption{The finite size scaling of the minority win probability ($w$) and its complimentary function $1-w$ (probability of majority win) are shown for the Ising model in a  two-dimensional topology. 
The simulations were done for $T=1.03T_c$, the number of groups is $M=49$. The weight factors for each group are chosen from the actual numbers of the
electoral college votes from different states (see Fig. \ref{elec_clg}).
The finite size scaling gives an exponent value $\alpha_{Ising} \approx 0.7$ (see Eq. (\ref{scaling_form})). The corresponding values of the x-axis are calculated from the real data and are shown. The cases of 1876 and 2016 falls considerably out of the range where $w$ is non-zero. The inset shows the unscaled data.}
\label{collapse_2dising} 
\end{figure}

We measure $w$, the probability for minority win at different temperatures as a function of $\Delta$
that arises from the fluctuations in the system. For $T< T_c$, it vanishes with the system size rapidly, showing 
that below the critical point, the probability of minority win is essentially zero.  For $T> T_c$, however, we find that $w$ shows a data collapse  when plotted against 
$\Delta/N_v^{0.5}$  when $T\gg T_c$ (not shown).  This remains true (in the thermodynamic limit) until  $T\approx T_c$, where the  scaling behavior changes to a form  
\begin{equation}
w=f(\Delta/N_v^{\alpha_{Ising}}),
\label{scaling_form} 
\end{equation}
with $\alpha_{Ising} \simeq 0.7$.
This scaling is the sole result of coarse graining near $T_c$.
We show the data collapse for  $T=1.03T_c$ in  Fig. \ref{collapse_2dising}.  
 One may add that the same scaling form is retained when all the blocks have equal
weightages (not shown).  However,  a distribution of  the weights considerably  enhances the values of $w$.

As for comparison with real data,
   it is found that clearly the results indicate a nonzero
probability corresponding to the  $r_{\alpha_{Ising}}=\Delta/N_v^{\alpha_{Ising}}$ values for the  years 2000 and 1888 while those for 2016 and 1876 fall in the regime where $w$ obtained from the simulations gives almost zero probability. Before commenting on this 
we proceed to calculate $w$ for a more realistic model.

\subsubsection{Kinetic exchange model (KEM)}
One major drawback of the Ising model is the absence of a third state, which can represent non-participation and third party voting.
We focus on a kinetic exchange model of opinion formation \cite{BCS} which allows three different opinions $\pm 1$, 0 (so, $N_v<N$ here).  
Here the agents change their opinions based on binary interaction with another 
one.  This is similar
to previously explored kinetic exchange models \cite{lccc,sen2,sen3,soumya,Deff,HK,toscani}, but in most cases interaction brought 
opinions closer while  in this model  the interactions
 can either bring  opinions together or push them further apart, depending on the state of relation between the interacting agents. Particularly,
the opinion of the $i$th individual at time $t+1$ is given by $o_i(t+1)=o_i(t)+\mu_{ij}o_j(t)$, where $o_m\in{+1,0,-1}$ and $\mu_{ij}$ can be
$-1$ with probability $p$ and $+1$ otherwise. There is no summation over $j$ on the second term, as the interactions are between only two agents at a time. 
The opinions are bounded between the two limits $-1:+1$.
The bound is the only source of non-linearity in the model. The tuning parameter is the noise $p$, increase of which signifies enhanced
polarization/discord in the society. The behavior of this model, particularly in terms of a global order parameter $O(t)=|\sum\limits_io_i(t)|/N$
in the limits of large time and system size $N$, are well studied \cite{nuno1, nuno2, khaleque}. In the case where any agent can interact with any other (mean-field limit), 
it is known exactly that at noise below $p_c=1/4$ \cite{BCS}, the population forms a consensus with either positive or negative
opinion dominating. Recently, the model has been studied in finite dimensions also \cite{sudip}. The associated criticality 
is found to be Ising like in all cases up to three dimensions.

\begin{figure}
\includegraphics[width=9cm]{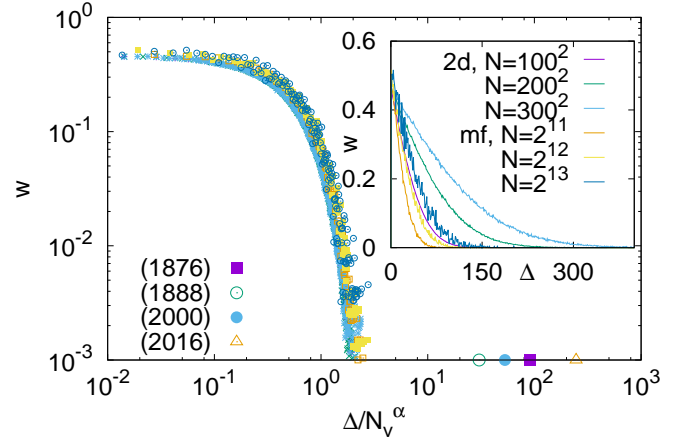}
\caption{The finite size scaling of the minority win probability ($w$) is shown for the interacting kinetic exchange
model for a fully connected or mean-field topology and a two-dimensional topology (data collapsing on two different curves). The simulations for the mean-field
graph were done for $p=0.27$, which is slightly above the critical value ($p_c=0.25$) for the transition to non-consensus.
The simulations for two-dimensional lattice were done for $p=0.7$, which is far from the critical point.
The scaling exponent value $\alpha_R=0.5$ is same as that found in the random assignment case. This highlights the non-existence
of spatial correlation in these two cases. As before, the scaled variables for the real data lie far away from
where the scaled function has become zero. The inset shows the unscaled data.}
\label{collapse_2d_mf_farct} 
\end{figure}
We begin by looking at the mean-field version of the model. The exact critical point  in the mean field case is known  to be $p_c=1/4$,  making it a simpler option to start with. We proceed as before, now with $M=50$ groups and retaining the unequal weightages to the groups. The noise value is kept fixed at $p=0.27$. The process of calculating  $w$ is exactly the same as in the case of Ising model except that $m(t)$ is now defined as $m(t) = \sum\limits_i o_i(t)$. For the finite size scaling, we have to use the number of agents ($N_v$) having non-zero opinion values, because for the real data we only compare with the total number of votes cast and not the voting age population (for the Ising model, of course, $N=N_v$). The finite size scaling of the minority win probabilities gives
$\alpha_{MF}=0.5$ (see Fig. \ref{collapse_2d_mf_farct}), which is what we expect for the cases with no spatial correlations. The comparison with real data is similar to the random case, as $w$ is negligible
for the real values of the argument $\Delta/N_v^{\alpha_{MF}}$. Hence we conclude, spatial fluctuations is an important feature in the model
and it is imperative to include it in low  dimensions.

\begin{figure}
\includegraphics[width=8cm]{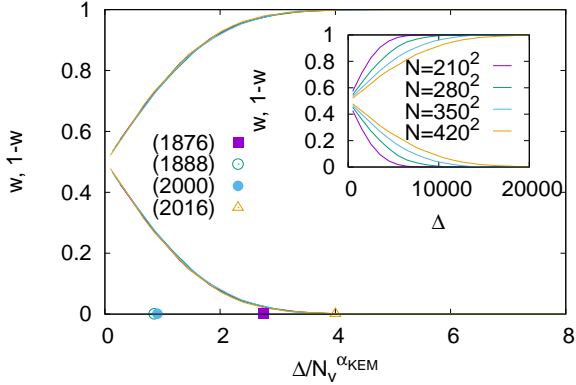}
\caption{The same as before for the two-dimensional version of the interacting 
model. The simulations were done at $p=0.12$. The value of the scaling
exponent $\alpha_{KEM}\approx 0.72$ imply a non-trivial spatial correlation and also similar to what is seen ($0.7$) in 
the Ising model (Fig. \ref{collapse_2dising}). The scaled variables for the real data are accommodated better than the Ising model,
and include all the four cases of the minority win, 2016 being right on the boundary. 
One reason could be scaling by the population size ($N_v= (1-g(p))N$; with $g(0.12)\approx 0.2$) only having non-zero opinion values.
 The inset shows the unscaled data.}
\label{minwin_2dbcs_p0.12}
\end{figure}
 While attempting to simulate the KEM in  two dimensions,  one faces the problem that $p_c$ 
 is known only numerically here, $p_c \approx 0.13$ \cite{sudip}.  This value is obtained with 
helical boundary conditions, and for periodic boundary conditions that we have followed throughout 
here, the critical point is $p_c \approx 0.11$ \cite{pvt}. Simulating the model far from $p_c$ will not retain any effect of the 
spatial fluctuation, and we will get back the mean-field type exponents (see Fig. \ref{collapse_2d_mf_farct}).
Therefore, we  simulate  the model at $p=0.12$, which is still above the critical point, to investigate the behavior of $w$ near criticality.  In the simulations, we keep the same structure as for the two dimensional Ising model and group the spins spatially into 49 groups and put different weight factor
to the groups drawn from the actual list of electoral college vote numbers of different states (and also checked the scaling for a uniform distribution in $[0:1]$).

Once again, $w$ vanishes for  $p<p_c$.
At high values of $p$, the minority winning probability shows scaling with 
$\Delta/N_v^{0.5}$. But at $p=0.12$, which is very close but just above the critical point,  one gets again a nontrivial result:  $w$ varies as    $\Delta/N_v^{\alpha_{KEM}}$ with $\alpha_{KEM}  \approx  0.72$
 (see Fig. \ref{minwin_2dbcs_p0.12}).   
Now we find that not only the points for two  elections' (1888 and 2000) values of  $r_{\alpha_{KEM}}= \Delta/N_v^{\alpha_{KEM}}$  are well within the  region where $w$ is nonzero, but also 1876 and 2016
are now within the range, predicting finite probability for minority win in all those cases (2016 being right at the boundary).
 The predictions are better than the two choice case of the Ising model, as we can now include 
non-participating voters in the model. 

\begin{figure}
\includegraphics[width=8cm]{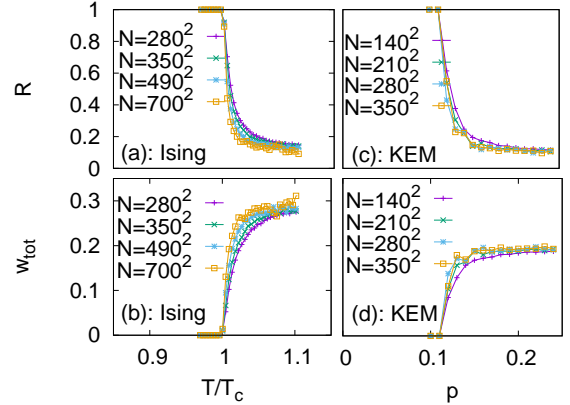}
\caption{Plot of the (a) relative mutual information
defined in Eq. (\ref{eq:rmi}) and (b) the total probability of the 
minority candidate win as a function of the noise $T/T_c$ for the Ising model.  Below the critical noise
$R$ is unity, implying full transmission of information and it decays sharply beyond 
that tending towards zero, which is the limit for complete loss of information.
The same are shown for the KEM in (c) \& (d) as a function of $p$ (not scaled by $p_c$, since it is not exactly known).
}
\label{ising_totprob}
\end{figure}

\subsection{Coarse-graining as loss of information}
The phenomenon of the minority winning is a direct consequence of loss of information due to
coarse graining of the system. Here we attempt a formal measure of it.

An important question is the overall probability of a minority win $w_{tot}$ when averaged over the individual 
probabilities for all values of $\Delta$. This is a measure of how far the result of election can be 
distorted due to coarse-graining and hence is a measure of the information loss.  We find that this probability 
typically remains zero below the critical point when
one of the opinions dominate, rapidly 
rises beyond it  and remains almost constant  as the noise increases above the critical point 
(see Fig. \ref{ising_totprob}).  
If one takes an average of $w$ for the Ising model over a reasonable range of noise (say from $T=0 $ to $2T_c$)  
 one gets  roughly  12\% cases 
where minority win is possible while the prediction from the kinetic exchange model is about 10\%. 
The percentage of the US presidential elections where minority win has occurred till date is $\sim 7$\% and thus one may expect more 
such events in future according to the present results.

A more formal measure for the information loss can be formulated with the help of classical information theory. 
It is a common practice \cite{hemmen} in various biological examples, such as neural firing, to measure the the relative 
information communicated between an input signal (stimuli) and the corresponding output (neural excitation).  
A more general notion is to consider the transmission of the signal through a noisy channel \cite{shannon}. 

In the present case, 
the input series is the time sequence of the sign of majority in the total population in each time step and the
output series is the sign of the majority of the corresponding coarse-grained system at those times. Ideally 
the two series should match at every instant, but the minority win cases are the deviations from that situation. 
Here we represent the positive sign majority by 1 and negative sign majority by 0. 
Let the input series (majority sign of $N$ agents) be denoted by $\mathcal{N}$ and the output series (majority sign 
of $M$ electoral college votes) be denoted by $\mathcal{M}$.  Then the mutual information ($I$) transmitted from the input
to the output is given by
\begin{equation}
 I(\mathcal{N},\mathcal{M})=H(\mathcal{N})+H(\mathcal{M})-H(\mathcal{N},\mathcal{M}),
\end{equation}
where
\begin{eqnarray}
H(X) &=& -\sum\limits_{i\in\{0,1\}}p(X=i)\log p(X=i),\nonumber \\
H(X,Y) &=& -\sum\limits_{i,j\in\{0,1\}} p(X=i \wedge Y=j)\log p(X=i \wedge Y=j) \nonumber
\end{eqnarray}
with $p(X=i)$ being the probability of input being $i$ and so on \cite{note1}.  
Then the relative mutual information ($R$) is calculated as \cite{szcz, agni}
\begin{eqnarray}
R(\mathcal{N},\mathcal{M})&=&\frac{H(\mathcal{N})+H(\mathcal{M})-H(\mathcal{N},\mathcal{M})}{[H(\mathcal{N})+H(\mathcal{M})]/2} 
\label{eq:rmi}
\end{eqnarray}
which is a measure of the reduction in uncertainty of the input, given the knowledge about
the realization of the output, relative to the average uncertainty of the input and output.
In Fig. \ref{ising_totprob}(a,c) the $R$ values are plotted against the external noise $T/T_c$ and $p$ ($p$ is not scaled by $p_c$ 
since it is not known exactly)
respectively for the Ising model and the KEM \cite{note}. Below the critical point, the $R$ is unity, 
implying that the output is fully predictable given the input. Beyond the critical point, $R$
decreases sharply and tends to its other limiting value zero, where the input and output are completely independent (all information is lost).
 This signifies the loss of information or unpredictability of the coarse-grained
majority in presence of relatively high noise.

\section{Discussions and conclusion}
The partitioning of a voting  population into smaller groups and then placing `winner takes all' rule in those groups
can result in a situation where the candidate with less number of total votes can win the overall election. A recent
example being the Presidential election in the US in 2016. We attempt here to capture the phenomena and and characterize
the universal scaling of the probabilities of such events through simple, discrete models of opinion dynamics (see Fig. \ref{snapshot} for
a snap-shot of a configuration where minority candidate wins). We have divided our model systems in equal parts, but have
placed different weight-factors in each part (state) to account for the unequal representations from each state.

We  have provided an estimate for the probabilities of minority win for given values of $\Delta$, the difference between the votes obtained
by the two major candidates. 
In very noisy systems $\Delta$ is small. For such a case, the 
 resultant scaling gives $\Delta \propto  N_v^{1/2}$ as one would expect in a completely 
random scenario (see sec. \ref{random_assign}). However,  such large values of noise are not observed in reality, as $N_v$, though large ($\sim 10^8$), 
is not infinite. Hence the results  well above the critical point cannot explain 
any of the real data (Fig. \ref{collapse_2d_mf_farct}).

On the other hand, fluctuations near the critical point in two-dimensional models (both Ising and KEM), significantly affect
the scaling, giving $\Delta \propto N_v^{\alpha}$ with $\alpha\approx 0.7$ for both the models.  Therefore, the finite size 
scaling is found to be a universal feature of the models with Ising symmetry. Furthermore, the weightage distributions of the blocks do not change the exponent values. We have checked this for both the actual distribution of the electoral college votes from different states in the US and also a uniform distribution in $[0:1]$.
The scaling  can, therefore,  be attributed to the process of coarse graining
near the critical point. 
 
 Also note that a common trend (see Fig. \ref{US_alldata}) for relatively high values of $\Delta$ is a linear 
variation with $N_v$. This can be understood from the models as follow:
$|\Delta|/N_v$ is basically the order parameter for the Ising model and the KEM. Clearly, when the order parameter value is large, i.e. the system is
deep into the ordered state, the order parameter is generally independent of the system size $N_v$, which implies $|\Delta| \propto N_v$ in this region. 
The proportionality also implies that the `information' present in the original system gets fully transmitted.  Indeed, the phenomenon of the minority candidate winning can be cast as loss of information due to coarse graining. This is measured in Fig. \ref{ising_totprob}
in terms of the total winning probability of the minority candidate winning $w_{tot}$ and the relative mutual information $R$ between the original 
and the coarse grained systems. In both cases, the critical point is where the system sharply transits into drastic reduction of information content.

In conclusion, the  counter-intuitive but not very rare results of the US presidential elections where the minority candidate won, can be explained 
 in terms of the non-consensus region of interacting models where noise drives a phase transition and yet does not take the system 
too far beyond the critical point. The crucial link is through the nontrivial 
functional form of the minority win probability $w(\Delta)$ varying as a function of the popular vote margin divided by the scaled voter number
$\Delta/N_v^{\alpha}$, with $0.5<\alpha<1$. The effect of non-participation in voting is critical in explaining the real data. The prediction of
finite probabilities for minority win implies that such effects should be taken care of in pre-election surveys while predicting the overall winning probability,
 particularly when the lead is small.

SB and PS acknowledge Alexander von Humboldt foundation and Council of Scientific \& Industrial Research, Government of India, respectively, for financial supports. 
The authors also thank Serge Galam and Dietrich Stauffer for their comments and discussions.


\begin{thebibliography}{99}
\bibitem{book1}
B. K. Chakrabarti, A. Chakraborti, A. Chatterjee, 
Editors, {\it Econophysics and sociophysics: Trends and perspectives},
Wiley-VCH, Weinheim, 2006.

\bibitem{rmp} C. Castellano, S. Fortunato, V. Loreto, {\it Statistical physics of social dynamics}, Rev. Mod. Phys. {\bf 81}, 591 (2009).

\bibitem{galam_book}
S. Galam, {\it Sociophysics: A physicist`s modeling of psycho-political phenomena}, Springer, 2012.

\bibitem{socio_book} P. Sen, B. K. Chakrabarti, {\it Sociophysics an Introduction}, Oxford University 
Press, 2013.


\bibitem{travieso} G. Travieso,  L. da Fontura Costa, 
{\it The spread of opinions and proportional voting},
Phys. Rev. E, {\bf 74}, 036112 (2006).



\bibitem{fort-cast07}
 S. Fortunato,  C. Castellano, 
{\it Scaling and universality in proportional elections},
Phys. Rev. Lett. {\bf  99}, 138701 (2007).




\bibitem{alves}
S. G. Alves, N. M. Oliveira Neto, M- L.  Martins, 
{\it Electoral surveys' influence on the voting
  processes: A cellular automata model},
Physica A {\bf  316},  601 (2002).



\bibitem{chatt_elec}
 A. Chatterjee, M. Mitrovi\'c, S. Fortunato, 
{\it Universality in voting behavior: an empirical analysis},
Sci. Rep. {\bf  3},   1049 (2013).

\bibitem{sinha}
S. Sinha, R. K. Pan, {\it How a ``Hit" is born: The emergence of popularity from the dynamics of collective choice}, in Ref. \cite{book1}.

\bibitem{note}
{We are not considering the special situation of the election in 1824, but keeping the 1876 case, although disputed.}
\bibitem{database} https://www.270towin.com
\bibitem{skma} S. K. Ma, {\it Modern theory of critical phenomena}, Westview Press, 2000.
\bibitem{galam86}
S. Galam, {\it Majority rule, hierarchical structure and democratic totalism: A statistical approach}, J. Math. Psychol. {\bf 30}, 426 (1986).
\bibitem{galam2002}
S. Galam,  {\it Minority opinion spreading in random geometry}, Eur. Phys. J. B,  {\bf 25},  403 (2002).

\bibitem{krapv}
P. L. Krapivsky, S. Redner, {\it Dynamics of majority rule in two-state interacting spin system}, Phys. Rev. Lett. {\bf 90}, 238701 (2003).
\bibitem{nard}
C. Nardini, B. Kozma, A. Barrat, {\it Who's taking first? Consensus or lack thereof in coevolving opinion formation model}, Phys. Rev. Lett. {\bf 100}, 158701 (2008)

 



\bibitem{x1}
M. G. Neubauer, M. Schilling, J. Zeitlin, {\it Exploring unpopular presidential elections}, arxiv:1206.2683.

\bibitem{x2}
R. S. Erikson, K. Sigman, {\it A simple stochastic model for close US presidenial elections}, 
(http://www.columbia.edu/~ks20/Erik-sig-academic.pdf)

\bibitem{gracia}
J. Fernandez-Gracia, K. Suchecki, J. Ramasco,
M. S. Miguel, V. M. Eguiluz,
{\it Is the voter model is a model for voters?},
Phys. Rev. Lett. {\bf 112}, 158701 (2014).


\bibitem{r1}
M. J. Hinich, R. Mickelsen, P. C. Ordeshook, {\it The electoral college vs. a direct vote: policy bias,
reversals and indeterminate outcomes}, J. Math. Sociol. {\bf 4}, 3 (1975).

\bibitem{r2}
C. Beisbart, L. Bovens, {\it Minimizing the threat of a positive majority deficit in
two-tier voting systems with equipopulous unis}, Public Choice {\bf 154}, 75 (2013).




\bibitem{galamx3}
S. Galam, {\it The Trump phenomenon, an explanation from sociophysics}, arxiv:1609.03933 (2016).



\bibitem{BCS} S. Biswas, A. Chatterjee, P. Sen, {\it Disorder induced phase transition in kinetic models of opinion dynamics}, Physica A {\bf 391}, 3257 (2012).

\bibitem{voter}
P. Clifford, A. W. Sudbury, {\it A model for spatial conflict}, Biometrika {\bf 60}, 581 (1973).

\bibitem{lccc}
 M. Lallouache, A. S.  Chakrabarti, A.  Chakraborti, B. K.  Chakrabarti,
{\it Opinion formation in kinetic exchange models: Spontaneous symmetry-breaking transition},
Phys. Rev. E {\bf 82}, 056112 (2010).

\bibitem{sen2}
 P. Sen, {\it Phase transitions in a two parameter model of opinion dynamics with random kinetic exchanges}, Phys. Rev. E {\bf  83}, 16108 (2011).

\bibitem{sen3}
 P. Sen,
{\it Nonconservative kinetic exchange model of opinion dynamics with randomness and bounded confidence},
Phys. Rev. E {\bf 86}, 016115 (2012).

\bibitem{soumya} S. Biswas, 
{\it Mean field solutions of kinetic exchange opinion models},
Phys. Rev. E {\bf 84}, 056106 (2011).

\bibitem{Deff}
 G. Deffuant, D. Neau, F. Amblard, G.  Weisbuch,  {\it Mixing beliefs
among interacting agents}, Adv. Complex Sys. {\bf 3}, 87 (2000).

\bibitem {HK}
R. Hegselmann, U.  Krause, {\it Opinion dynamics and bounded confidence models, analysis, and simulation},
J. Art. Soc.
Soc. Simul. {\bf 5}, 2 (2002).
\bibitem{toscani} G. Toscani, 
{\it Kinetic models of opinion formation},
Commun.  Math. Sci. {\bf 4}, 481 (2006).



\bibitem{nuno1}
N. Crokidakis, C. Anteneodo, {\it Role of conviction in nonequlibrium models of opinion formation}, Phys. Rev. E {\bf 86}, 061127 (2012).

\bibitem{nuno2}
N. Crokidakis, V. H. Blanco, C. Anteneodo, {\it Impact of contrarians and intransigents in a kinetic model of opinion dynamics}, Phys. Rev. E {\bf 89}, 013310 (2014).

\bibitem{khaleque}
A. Khaleque, P. Sen, {\it Damage spreading transition in an opinion dynamics model}, Physica A {\bf 413}, 599 (2014).

 





\bibitem{sudip}
S. Mukherjee, A. Chatterjee,  {\it Disorder-induced transition in an opinion dynamics model: Results in two and three dimensions}, Phys. Rev. E {\bf 94}, 062317 (2016).

\bibitem{pvt}
N. Crokidakis, private communication.

\bibitem{hemmen}
J. L. van Hemmen, T. Sejnowski, {\it 23 Problems in Systems Neurosciences}, Oxford University Press, UK (2006).

\bibitem{shannon}
C. E. Shannon, {\it A Mathematical Theory of Communication}, The Bell System Technical Journal {\bf 27}, 379 (1948).

\bibitem{note1}
While calculating the $R$ for KEM, for the time series below the critical point, the signs of all the agents are  flipped
with some probability, in order to avoid getting undetermined denominator in Eq. (\ref{eq:rmi}). The problem does not appear in the Ising 
model, since we use cluster algorithm, where the largest cluster is flipped anyway.
\bibitem{szcz}
J. Szczepanski, M. Arnold, E. Wajnryb, J. M. Amig\'{o}, M. V. Sanchez-Vives, {\it Mutual information and redundancy in spontaneous communication between cortical neurons}, Biological Cybernetics {\bf 104}, 161 (2011).

\bibitem{agni}
A. Pregowska, J. Szczepanski, E. Wajnryb, {\it Mutual information against correlations in binary communication channels}, BMC Neuroscience {\bf 16}, 32 (2015).



\end{thebibliography}
\end{document}